\documentclass[aps,pra,twocolumn,10pt, showpacs,superscriptaddress]{revtex4-2} 
\usepackage[utf8]{inputenc}
\usepackage{physics}
\usepackage{graphicx}
\usepackage{epstopdf}
\usepackage{setspace}
\usepackage{bbold}
\usepackage{comment}
\usepackage{color}
\usepackage{soul}
\usepackage[abs]{overpic}
\usepackage{amsmath}
\usepackage{lipsum}
\usepackage{placeins}
\usepackage[abs]{overpic}
\usepackage{float}
\usepackage[export]{adjustbox}
\usepackage{amsmath}
\usepackage{nicefrac}
\usepackage{cleveref}
\usepackage{layouts}
\usepackage[font=small, labelsep=period,
   justification=Justified,
   format=plain]{caption} % 'format=plain' avoids hanging indentation
\usepackage{ragged2e}

\graphicspath{ {./images/} }
\begin{document}

\title{Enhanced synchronization with proportional coupling in Kuramoto oscillator networks under a fixed coupling budget}
\author{Amit Pando}
\affiliation{Department of Physics of Complex Systems, Weizmann Institute of Science, Rehovot 7610001, Israel}
\author{Eran Bernstein}
\affiliation{Department of Physics of Complex Systems, Weizmann Institute of Science, Rehovot 7610001, Israel}
\author{Tomer Hacohen}
\affiliation{Department of Physics of Complex Systems, Weizmann Institute of Science, Rehovot 7610001, Israel}
\author{Nathan Vigne}
\affiliation{Department of Applied Physics, Yale University, New Haven, CT, USA }
\author{Hui Cao}
\affiliation{Department of Applied Physics, Yale University, New Haven, CT, USA }
\author{Oren Raz}
\affiliation{Department of Physics of Complex Systems, Weizmann Institute of Science, Rehovot 7610001, Israel}
\author{Asher Friesem}
\affiliation{Department of Physics of Complex Systems, Weizmann Institute of Science, Rehovot 7610001, Israel}
\author{Nir Davidson}
\affiliation{Department of Physics of Complex Systems, Weizmann Institute of Science, Rehovot 7610001, Israel}
\begin{abstract}

We propose an explicit coupling scheme for enhancing the synchronization of Kuramoto oscillator networks under a fixed coupling budget. We show that by scaling the interaction strength between oscillators according to their frequency detuning, synchronization is enhanced. The coupling scheme produces a crossover from a smooth Kuramoto onset to an abrupt, hysteretic synchronization transition as a single allocation parameter is varied. Our work offers a general route to efficient synchronization in engineered Kuramoto networks and provides insight into their critical behavior.
\end{abstract}
\maketitle

\textit{Introduction -} Many different physical, chemical and biological systems can be modeled as networks of coupled oscillators \cite{pikovsky2001synchronization,strogatz2012sync}. These networks are often heterogeneous, with each oscillator having its own natural frequency, yet they can synchronize to oscillate with a common frequency and phase. The Kuramoto model can accurately describe the dynamics of these networks and their transition to a synchronized state, and was successfully applied to analyze swarms of fireflies, arrays of coupled Josephson junctions, arrays of coupled lasers, and even human crowds \cite{RevModPhys.77.137,shahal2020synchronization,kuramoto2003chemical,trees2005synchronization,PhysRevLett.133.113803}. Synchronization is induced by mutual coupling between the different oscillators, and different coupling schemes affect the existence, dynamics, and stability of the synchronized state.

In many applications synchronization must be achieved with constrained coupling strengths and connectivity \cite{li2017optimizing,pinto2015optimal,kelly2011topology,ye2026disorder,barioni2025interpretable}. For example,  the coupling strength must be limited to prevent chaotic behavior of coupled lasers  \cite{donati2012chaos}, and the possible connectivity  is limited for chip-based systems such as Josephson junction arrays\cite{bravyi2022future}. Such constraints raise a fundamental question of how a fixed coupling resource should be allocated to improve synchronization in oscillator networks with a coupling budget (total coupling strength). 

In this work we employ a coupling heuristic, rather than using computational  optimization algorithms \cite{brede2008locals,tanaka2008optimal,nishikawa2006synchronization,li2017optimizing,mikaberidze2025emergent,brede2018competitive}, which exploits specific information about oscillators networks to significantly improve synchronization under a fixed coupling budget constraint. We enhance synchronization with a physically motivated and mathematically explicit coupling scheme, that depends only on measured natural frequencies of the oscillators. Specifically, the oscillators are coupled with coupling strengths that are proportional to their natural frequency difference. Previous research has shown that such proportional (or frequency-weighted) coupling scheme can change the critical behavior of the network \cite{leyva2013explosive}. We show that when subject to a coupling budget constraint, proportional coupling enhances the synchronization of the system compared to uniform all-to-all (or mean-field) coupling. We introduce an additional parameter to modulate the distribution of coupling between weakly detuned and far detuned oscillators. We show that in the presence of a coupling budget constraint, the effects of this parameter are non-monotonic, unlike those shown in previous works \cite{leyva2013explosive}. Tuning this parameter enables enhanced synchronization at any coupling magnitude and to fully synchronize the network at coupling strengths significantly smaller than for the well-studied uniform all-to-all coupling scheme.

\section{Proportional coupling}
The Kuramoto model \cite{RevModPhys.77.137} is typically described by
\begin{equation} \label{kuramoto_eqn}
    \dot{\theta_i} = \Omega_i + \sum_{j=1}^{N}K_{ij}\sin(\theta_j-\theta_i)
\end{equation}
where $\theta_i$ and $\Omega_i$ are the phase and intrinsic natural frequency of the i-th oscillator, $K_{ij}$ is the coupling between oscillators i and j, and time-dependent fluctuations are neglected. A well-studied paradigmatic case of the Kuramoto model involves a uniform all-to-all (mean-field) coupling scheme, where $K_{ij}=K/N$ \cite{RevModPhys.77.137}. Then, given some distribution of natural frequencies, $g(\Omega)$, it is necessary to determine whether and how well the network of oscillators will synchronize. Typically, the distribution of frequencies is symmetric about a central frequency, so we can move to a rotating frame where the average frequency $\ev{\Omega_i}=0$.  The usual approach involves introducing the phase-locking order parameter
\begin{equation}\label{eqn_order_param}
  re^{i\psi} = \frac{1}{N}\sum_j e^{i\theta_j}  
\end{equation}
where $r$ indicates the degree of phase-locking (the length of the average phasor) and $\psi$ the average phase. In the standard analytical solution, $r(K)$ undergoes a second-order phase transition in the $N\rightarrow \infty$ limit, with a critical point at $K = K_c \equiv \frac{2}{\pi g(0)} $ \cite{RevModPhys.77.137} .

We now consider how to improve synchronization when operating under a fixed coupling budget \cite{mikaberidze2025emergent}, of:
\begin{equation}\label{eqn_budget}
\frac{1}{N}\sum_{i,j}K_{ij} = K 
\end{equation}
We note that with uniform coupling, the budget cost for coupling oscillators of similar frequencies is the same as that for coupling oscillators of very different frequencies. Intuitively, such coupling is not efficient because oscillators with similar natural frequencies could be synchronized with weaker coupling between them, and free a larger portion of the coupling budget to couple more detuned oscillators. Previous studies of numerically optimized networks under a fixed budget have found preference for strong coupling between oscillators with large natural frequency difference \cite{mikaberidze2025emergent,tanaka2008optimal}. We thus surmise that synchronization can be improved by resorting to proportional coupling, defined by
\begin{equation}\label{eqn_prop}
    K_{ij} =\frac{K}{C(\{\Omega\})}\abs{\Omega_i-\Omega_j},
\end{equation}
where 
\begin{equation}
    C(\{\Omega\}) = \frac{1}{N}\sum_{i,j}\abs{\Omega_i-\Omega_j}
\end{equation}
is the realization-dependent budget normalization factor that ensures Eq. (\ref{eqn_budget}) is satisfied. Proportional (or frequency-weighted) coupling schemes \cite{leyva2013explosive}  were shown to induce changes in the critical behavior of the network. In our work, we found that with a budget constraint such changes are significantly different.

We numerically simulated a network of $N=512$ Kuramoto oscillators with normally distributed frequencies ($g(\Omega) \sim \mathcal{N}(0,\sigma^2)$) using both proportional and uniform coupling. The simulation starts from random initial conditions and uses a timestep size of  $dt=0.05$. A running average of the synchronization order parameter $r=1/\tau\int_{t-\tau}^{t}r(t)\dd{t}$ is calculated, and the simulation persists until $T=100$, where we validate that the value of $r$ has converged to a steady state. The average steady state order parameter $\bar{r}$ is evaluated by repeating this process for $20$ different random initial conditions and $100$ different disorder realizations. Additional numerical simulation details are presented in Appendix A.

\begin{figure}[h]
\centering
        \begin{overpic}
            [width=1\linewidth]{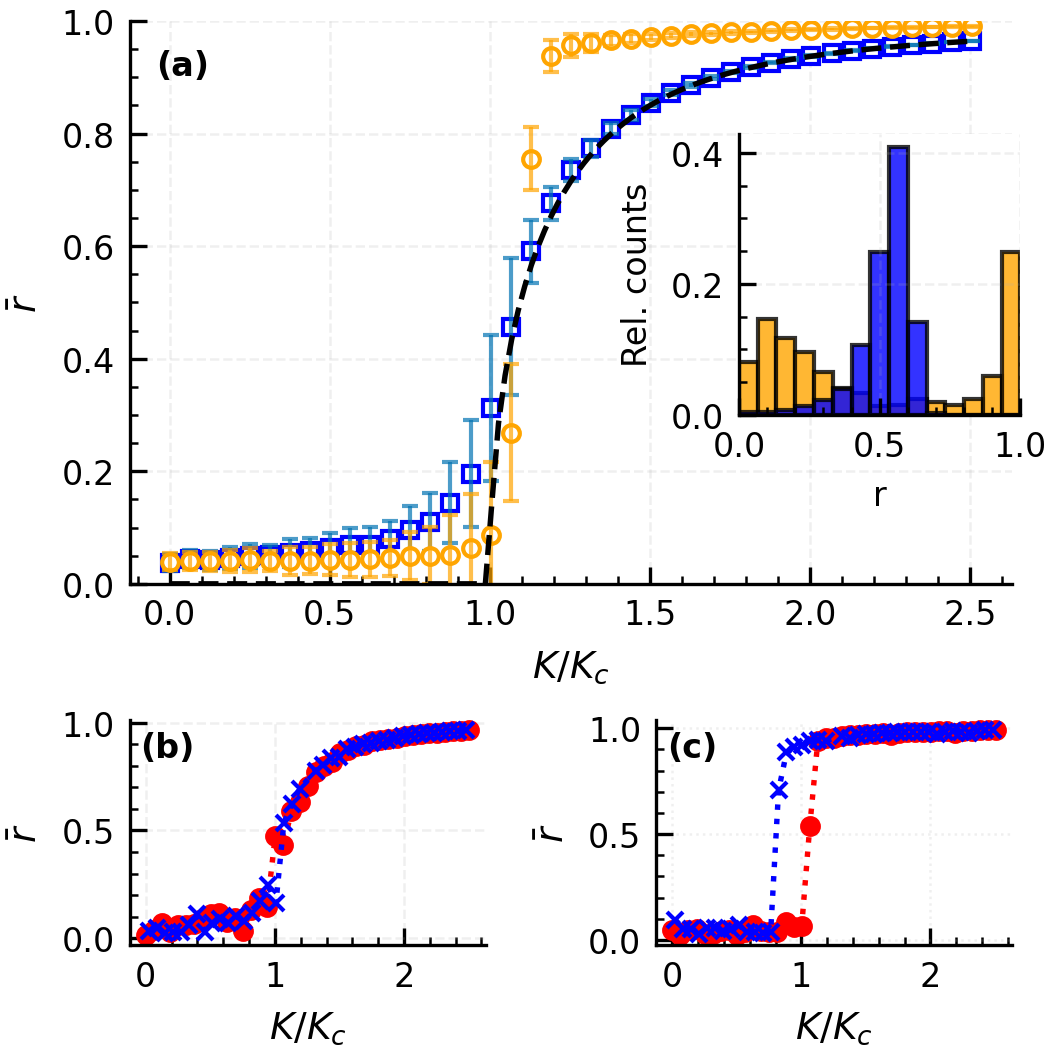}
            % \put(50,140){\large \textbf{(b)}}
        \end{overpic}%

    \caption{Synchronization with proportional and uniform coupling for a network of $N=512$ Kuramoto oscillators with normally distributed natural frequencies. \textbf{(a)}: Average order parameter $\bar{r}$ vs. normalized coupling strength $K/K_c$ with proportional coupling  (yellow circles) compared to that with uniform all-to-all coupling  (blue squares). Error bars represent standard deviation of the order parameter across random frequency detuning distribution realizations. Dashed black line shows the exact solution of Eq. \ref{kuramoto_eqn} with uniform coupling for $N\rightarrow \infty$ \cite{RevModPhys.77.137}. \textbf{Inset:} Histogram of the order parameter $r$ for $K/K_c=1.09$ showing a bimodal distribution in the case of proportional coupling in contrast to a unimodal distribution in the case of uniform coupling. \textbf{(b-c):} Hysteresis curves for uniform (b) and proportional (c) coupling. Red circles (blue crosses) correspond to the forward (backward) direction of adiabatically changing $K/K_c$.}
    \label{fig:prop_results}
\end{figure}

The simulation results are presented in Fig. \ref{fig:prop_results}. Figure \ref{fig:prop_results}(a) shows the simulated order parameter $\bar{r}$ as a function of the normalized coupling strength $K/K_c$. For small $K/K_c$ uniform coupling provides  somewhat better synchronization. However, around $K\approx 1.1K_c$ an abrupt transition in $\bar{r}$ occurs with proportional coupling, resulting in enhanced synchronization when compared to uniform coupling. 
The inset shows a histogram of $r$ for different realizations around the transition point. As evident, proportional coupling results in a bimodal distribution of $r$ values, indicating "all or nothing" behavior, as opposed to the narrow distribution around intermediate $r$ for uniform coupling. Specifically, if the coupling is sufficient to synchronize strongly detuned oscillators, then the entire network will synchronize. Figures \ref{fig:prop_results}(b) and \ref{fig:prop_results}(c) show the simulation results of synchronization order parameter $\bar{r}$  when adiabatically changing the coupling strength for the two coupling schemes. As evident, hysteresis behavior occurs for proportional coupling but not for uniform coupling. The hysteresis indicates that the synchronized state is metastable for proportional coupling even for $K<K_c$.

\begin{figure}[h]
\centering
    \begin{overpic}
        [width=1\linewidth]{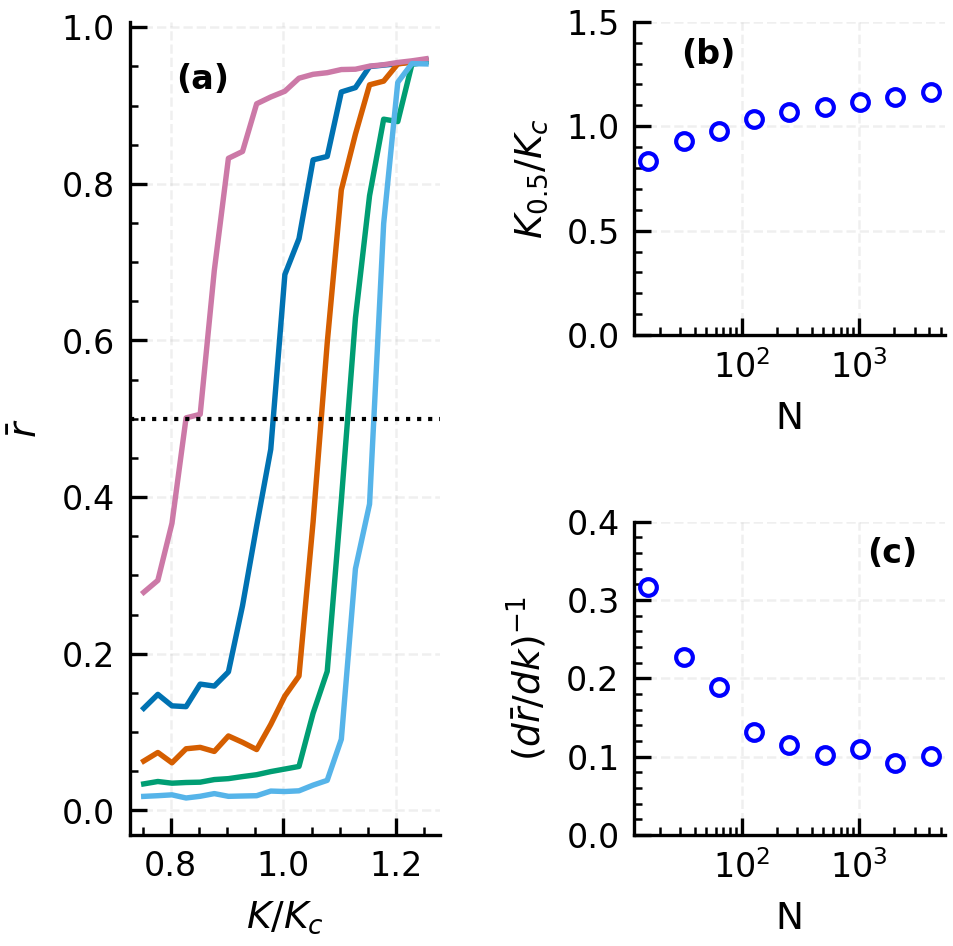}

    \end{overpic}%
    
    \caption{Effect of network size $N$ on synchronization with proportional coupling. \textbf{(a)} Synchronization order parameter $\bar{r}$ as a function of $K/K_c$ for  $N = 16$ (pink), $ 64$ (blue),  $256$ (orange), $1024$ (green), $4096$ (teal).\textbf{(b)}  Transition point $K(\bar{r}=0.5)/K_c$ (marked in a black dotted line in (a)) as function of $N$. \textbf{(c)} Inverse slope at the transition point as a function of $N$.}
    \label{fig:proportional_fss}
\end{figure}

We also investigated the effect of changing the size of the network on synchronization with proportional coupling. The results are presented in Fig. \ref{fig:proportional_fss}. Figure \ref{fig:proportional_fss}(a) shows the order parameter $\bar{r}$ as function of normalized coupling strength $K/K_c$ for network sizes ranging from $N=16$ to $N=4096$. As evident, there is an abrupt transition for each network size. We define a transition point as the value of $K$ for which $\bar{r}=0.5$ and find that it scales sub-linearly with $log(N)$, as shown in Fig. \ref{fig:proportional_fss}(b). The slope of the transition, shown in Fig. \ref{fig:proportional_fss}(c), also appears to saturate as $N$ is increased.

The appearance of abrupt transitions, hysteresis and bimodal behavior of the order parameter indicate a change of critical behavior induced when using the proportional coupling. These numerical results agree with semi-analytical results presented in the next section. Previous works with frequency correlated coupling schemes have observed the appearance of explosive synchronization transitions \cite{gomez2011explosive,vlasov2015explosive,zhang2013explosive,leyva2013explosive}. Our results show that the addition of a coupling budget constraint is compatible with the altered critical behavior.

We also examined synchronization with proportional coupling when the frequency distribution is not exactly known. To that end, we add random disorder to the coupling terms, such that $\varepsilon \equiv \text{Corr}(K_{ij},\abs{\Omega_i-\Omega_j})$ is varied in a controlled manner: $\varepsilon=1$ indicates proportional coupling, while $\varepsilon=0$ indicates random (positive) coupling. Additional details regarding the generation of a partially correlated coupling matrix can be found in Appendix B. The results, presented in Fig. \ref{fig:epsilon}, indicate that even partial correlation between the coupling terms and the detuning differences is sufficient to significantly enhance synchronization for $K>1.1K_c$. Our results imply that proportional coupling could be advantageous for synchronization even in applications where the oscillator frequencies cannot be determined accurately. The results also indicate that synchronization with random coupling is surprisingly similar to that of uniform coupling, where in both the coupling terms are not correlated with the detuning differences. This implies that correlation between the coupling and the frequency distribution of the networks plays a key role in enhancing synchronization.

\begin{figure}[h]
    \centering
        \includegraphics[width=1\linewidth]{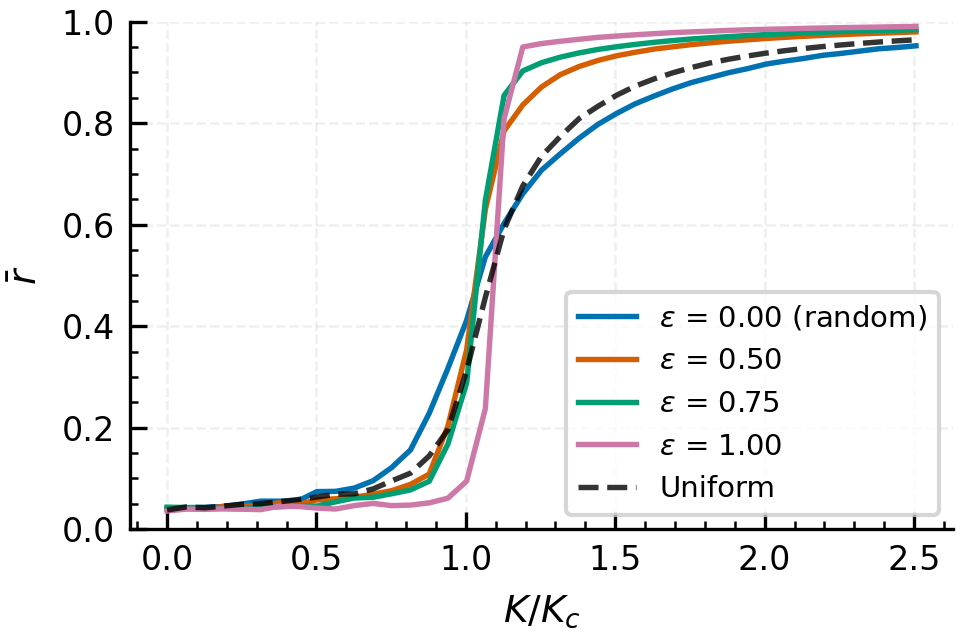}

    \caption{Synchronization order parameter $\bar{r}$ as a function of the normalized coupling strength $K/K_c$ with  partially correlated proportional coupling. The correlation between the coupling terms and the detuning differences is quantified by $\varepsilon=\text{Corr}(K_{ij},\abs{\Omega_i-\Omega_j})$. Uniform coupling (dashed line) is very similar to the random coupling case.}
    \label{fig:epsilon}
\end{figure}

\section{Sparse proportional coupling} 

\begin{figure}[h]
    \centering
    \includegraphics[width=1\linewidth]{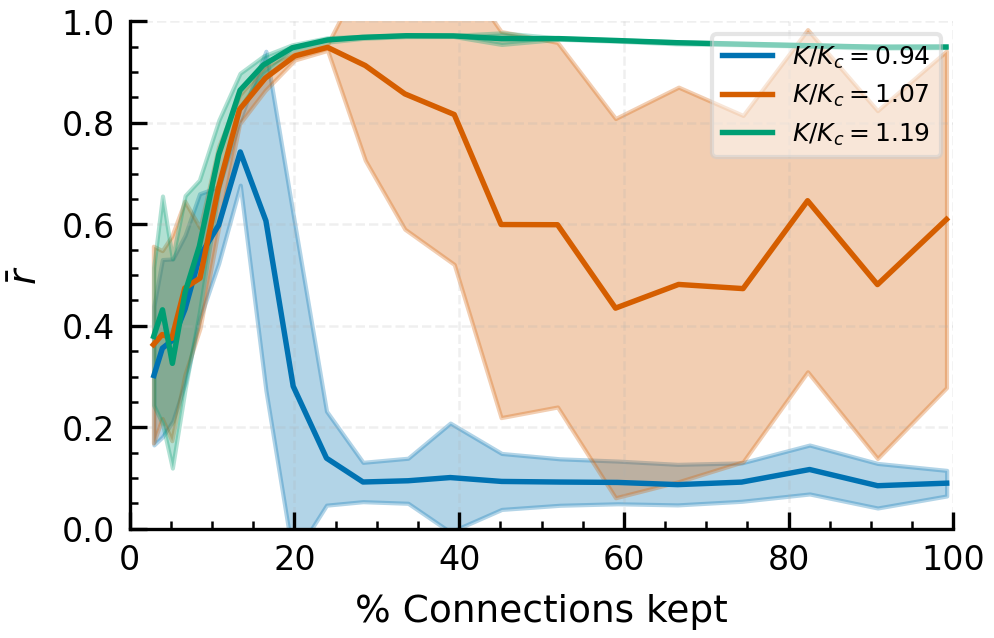}
    \caption{Synchronization order parameter $\bar{r}$ as function of remaining connections for sparse proportional coupling for $K/K_c=0.94$ (blue),$1.07$ (orange),$1.19$ (green). Shaded areas denote the standard deviation of the order parameter across random frequency detuning distribution realizations. Notably, disconnecting weak coupling terms does not diminish synchronization, and for marginal coupling magnitudes, can even improve it.}
    \label{fig_sparse_main}
\end{figure}

We also investigated the effects of network sparsity on proportional coupling. Specifically, we examined whether it is beneficial to correlate both the network structure and the magnitude of the coupling terms with the frequency distribution. To that end, we set the lowest coupling terms in $K_{ij}$ to zero and then renormalize the coupling matrix to the budget. The results of synchronization order parameter $\bar{r}$ as a function of remaining connections for sparse proportional coupling are presented in Fig. \ref{fig_sparse_main}. As evident, the network can synchronize when only $20\%$ of the coupling terms are retained. In other words, a large proportion of coupling terms (roughly $80\%$) do not effectively contribute to the synchronization of the network. Notably, for marginal coupling magnitudes ($K/K_c\sim1$), sparseness leads to improved (partial) synchronization. Recent works on computational optimization of Kuramoto networks have obtained sparse networks with similar structures \cite{mikaberidze2025emergent,ye2025optimal}.

\section{Generalized proportional coupling} In general, we can consider more complicated coupling of the form $K_{ij}=f(\abs{\Omega_i-\Omega_j})$. Here we resort to $K_{ij} \propto \abs{\Omega_i-\Omega_j}^p$, where $p$ is a tunable parameter, such that $p=0$ denotes uniform coupling, and $p=1$ denotes proportional coupling, while retaining the same budget constraint of Eq. (\ref{eqn_budget}).

\begin{figure}[h]
    \centering
    \includegraphics[width=1\linewidth]{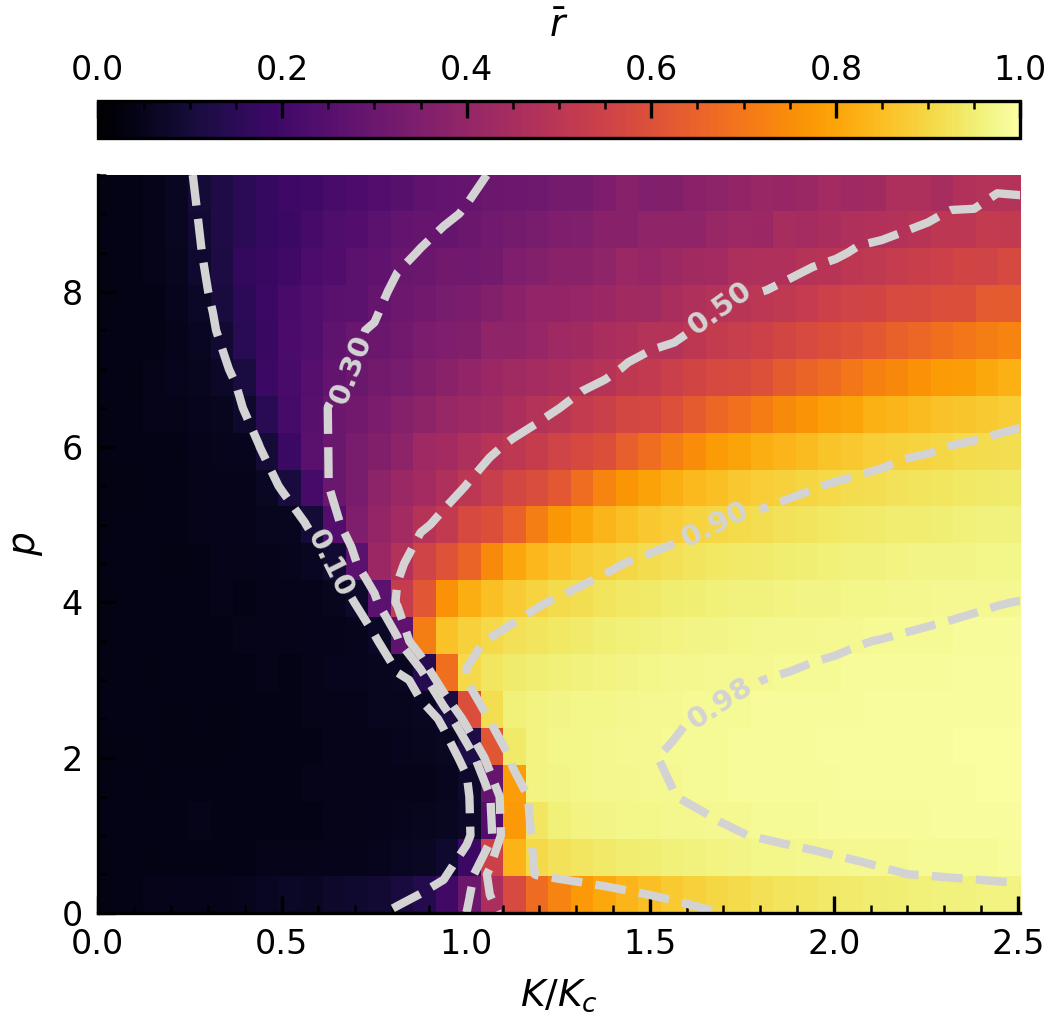}
    \caption{Synchronization of a network of $N = 512$ oscillators for $K_{ij}\propto\abs{\Omega_i-\Omega_j}^p$. An abrupt transition can be seen for $1<p<3$.}
    \label{fig:power_2d}
\end{figure}

Following the same simulation procedure as before, we determined the synchronization order parameter $\bar{r}$ as a function of the parameter $p$ and the normalized coupling strength $K/K_c$. The results are presented in Fig. \ref{fig:power_2d}. As evident, $\bar{r}$ varies non-monotonically with $p$. The sharpness of the synchronization transition appears to vary with $p$ as well. For $1<p<3$, $\bar{r}$ rises sharply around $K\sim K_c$, with the sharpest increase found for $p\approx 2$. For $p\ge4$, $\bar{r}$ increases continuously, with synchronization starting to emerge for values of  $K/K_c\ll 1$, but does not reach $\bar{r}\approx 1$ even at large $K/K_c$. Since generalized proportional coupling with large $p$ allocates the coupling budget primarily to the most detuned connections, the bulk of the oscillators in the network are weakly coupled and are unable to synchronize. We  obtained qualitatively similar results when considering uniform and Cauchy frequency distributions (Sec. II in \cite{Supplemental}) and Stuart-Landau oscillators (Sec. III of \cite{Supplemental}).

\begin{figure}[h]
    \centering
    \includegraphics[width=1\linewidth]{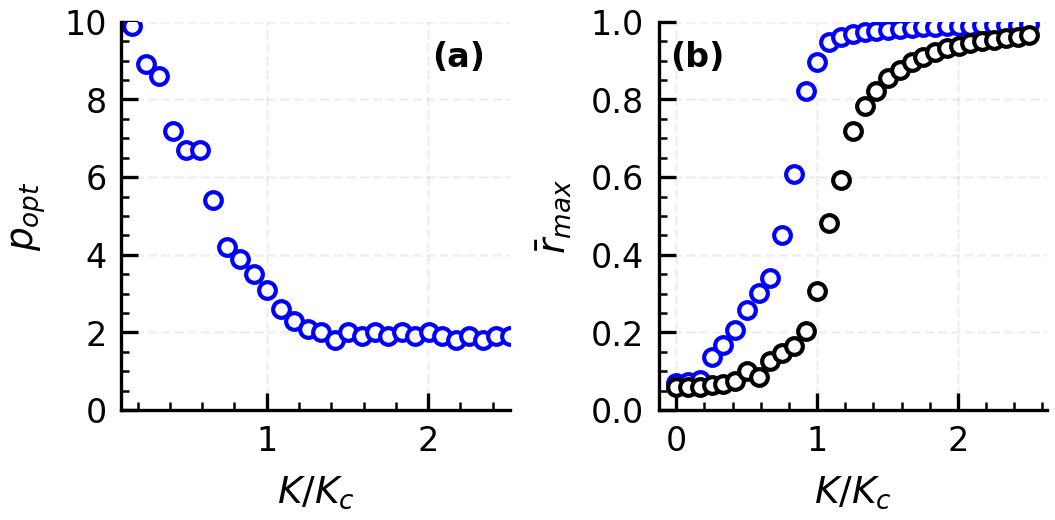}
    \caption{Enhanced synchronization with proportional coupling. \textbf{(a)} $p_{opt}$, the optimal value of $p$ that generates the maximal $\bar{r}$ for each $K/K_c$ for proportional coupling. \textbf{(b)} $\bar{r}_{max}$ obtained for each $p_{opt}$ of (a)  (blue circles) is higher than for uniform coupling  (black dots) for all values of  $K/K_c$.}
    \label{fig:optimal}
\end{figure}
\begin{figure}[h]
    \centering
    \includegraphics[width=1\linewidth]{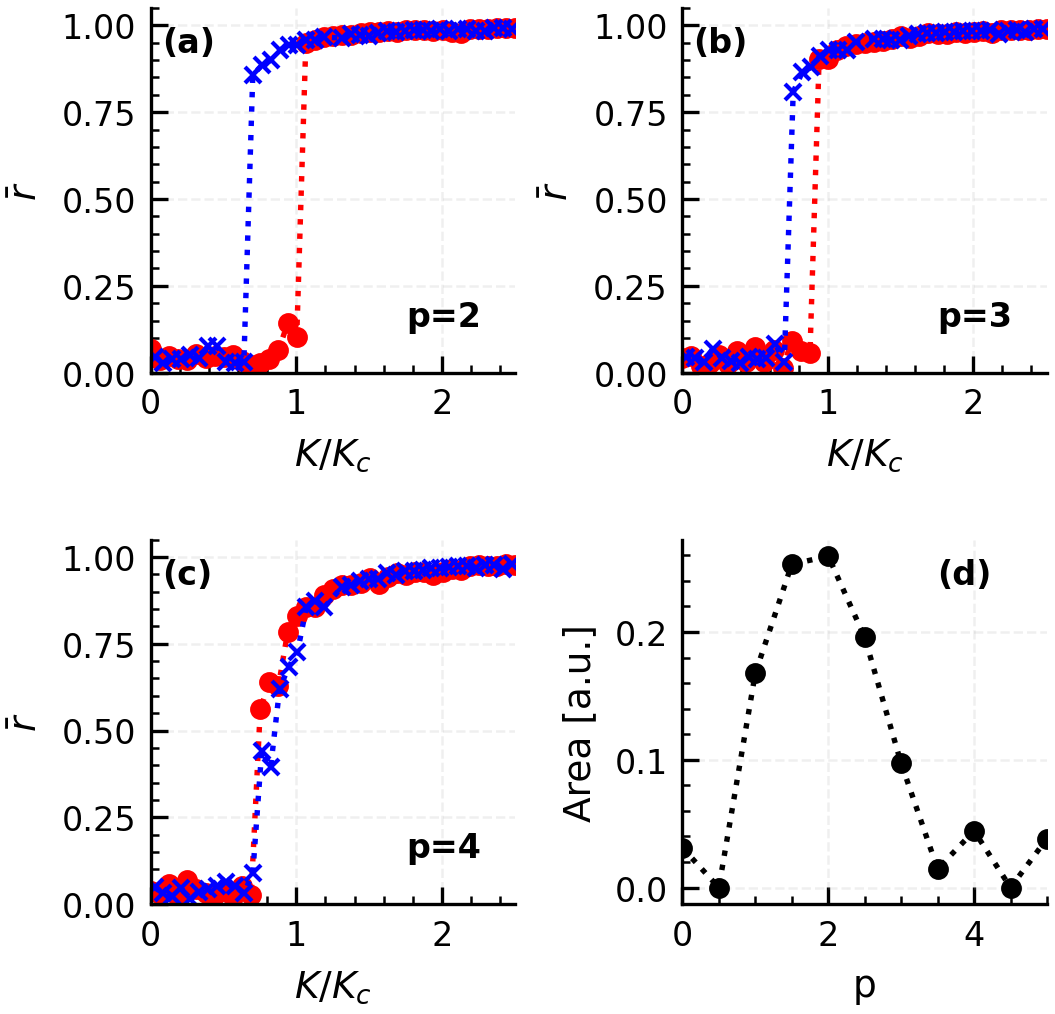}
    \caption{Hysteresis in the generalized proportional coupling scheme. \textbf{(a-c)} Hysteresis curves for $p=2,3,4$ respectively. Red circles (blue crosses) correspond to the forward (backward) direction of adiabatically changing $K/K_c$. \textbf{(d)} Area enclosed in the hysteresis curve as function of $p$ indicating most pronounced hysteresis around $p=2$.}
    \label{fig:hysteresis}
\end{figure}

It is evident from Fig. \ref{fig:power_2d} that for each $K$ there exists a value $p_{opt}$ that results in the largest $\bar{r}$. Figure \ref{fig:optimal}(a) shows a plot of $p_{opt}(K)$, and Fig. \ref{fig:optimal}(b) shows the corresponding $\bar{r}_{max}$ along with the results for uniform coupling. As evident, using $p_{opt}(K)$ the network can reach $\bar{r}\gtrsim 0.9$ for coupling magnitude as low as $K = 0.8K_c$. Also, $p\approx 2$ remains an optimal value for $K/K_c>1.1$.
Our synchronization results are comparable to those obtained by means of computational optimization algorithms \cite{mikaberidze2025emergent,ye2025optimal}.

Finally, we explored the effect of $p$ on synchronization when varying the coupling strength adiabatically. The results for different values of $p$ are presented in Fig. \ref{fig:hysteresis}. As for $p=1$ (shown in Fig. \ref{fig:prop_results}(c)), the results for larger $p$ in Figs. \ref{fig:hysteresis}(a-c) all show distinct, albeit different, hysteresis behavior. Figure \ref{fig:hysteresis}(d) shows that the area of the hysteresis loop varies non-monotonically with $p$, and is maximal for $p=2$, indicating that the synchronized state is metastable even for $K\sim 0.6K_c$. This is distinct from the results obtained without a coupling budget constraint, where it was shown that increasing $p$ will always increase the hysteresis loop area\cite{leyva2013explosive}.

\section{Analytical results}
We employ an analytical approach to capture the full behavior of Figs. \ref{fig:power_2d}-\ref{fig:hysteresis}\cite{xu2016synchronization}. Taking the continuum limit of the proportionally coupled Kuramoto equation we get
\begin{equation}\label{eqn_kuramoto_continuum}
\dot{\theta}= \Omega + \frac{K}{M_p}\int\abs{\Omega-\Omega'}^p g(\Omega')\ev{\sin(\theta'-\theta)}_{\Omega'}\dd\Omega',
\end{equation}
where $M_p$ is the weighted coupling normalization factor in the continuum limit
\begin{equation}
    M_p \equiv \iint \abs{\Omega-\Omega'}^p g(\Omega)g(\Omega')\dd\Omega\dd\Omega',
\end{equation}
and 
\begin{equation}\label{eqn_avg_sin}
    \ev{\sin(\theta'-\theta)}_{\Omega'} = \int_0^{2\pi} \rho(\vartheta,\Omega',t)\sin(\vartheta-\theta)\dd\vartheta,
\end{equation}
is the average interaction term for oscillators with natural frequency $\Omega'$ in terms of the oscillator density $\rho$\cite{RevModPhys.77.137}.

We define a frequency-resolved order parameter,
\begin{equation} \label{eqn_mf_eta}
    \eta(\Omega) \equiv \ev{e^{i\theta}}_{\Omega} = \int_0^{2\pi} e^{i\vartheta}\rho(\vartheta,\Omega,t)\dd\vartheta,
\end{equation}

and rewrite Eq. (\ref{eqn_avg_sin}) as
\begin{equation}
     \ev{\sin(\theta'-\theta)}_{\Omega'} = \ev{\Im[e^{-i\theta}e^{i\theta'}]}_{\Omega'} = \Im[e^{-i\theta}\eta(\Omega',t)],
\end{equation}
such that Eq. (\ref{eqn_kuramoto_continuum}) can be written as
\begin{equation}\label{eqn_kuramoto_continuum2}
    \dot{\theta} = \Omega + \frac{K}{M_p}\Im[e^{-i\theta}\int\abs{\Omega-\Omega'}^pg(\Omega')\eta(\Omega',t)\dd\Omega'].
\end{equation}
We define the integral in Eq. (\ref{eqn_kuramoto_continuum2}) as the complex field
\begin{equation}\label{eqn_mf_z}
    Z_p(\Omega) \equiv \int \abs{\Omega-\Omega'}^p g(\Omega')\eta(\Omega')\dd{\Omega'}\equiv A_p(\Omega)e^{i\phi_p(\Omega)}.
\end{equation}
This $Z_p$ quantifies the "mean-field" affecting an oscillator with frequency $\Omega$. 

We substitute Eq. (\ref{eqn_mf_z}) into the Kuramoto equation to get
\begin{equation}
    \dot{\theta}(\Omega) = \Omega + \frac{K}{M_p}A_p(\Omega)\sin(\phi_p(\Omega)-\theta), 
\end{equation}

For locked oscillators, $\dot{\theta}(\Omega)=0$, so
\begin{equation}
    \sin(\theta-\phi_p(\Omega)) = \frac{\Omega M_p}{K A_p(\Omega)}\equiv u_p(\Omega)
\end{equation}
\begin{equation}\label{eq_theta_locked}
    \theta(\Omega) = \phi_p(\Omega)+\arcsin{u_p(\Omega)}
\end{equation}
Substituting Eq. (\ref{eq_theta_locked}) into Eq. (\ref{eqn_mf_eta}), yields
\begin{equation}\label{eta_locked}
    \eta(\Omega) = e^{i\phi_p(\Omega)}[iu_p(\Omega)+\sqrt{1-u_p^2(\Omega)}].
\end{equation}
where $\abs{u_p(\Omega)}\le 1$ for locked oscillators.
For unlocked oscillators ($\abs{u_p(\Omega)}
\ge1$), we assume that the steady state phase density is inversely proportional to the magnitude of the angular velocity\cite{RevModPhys.77.137}, yielding
\begin{equation}\label{eta_unlocked}
    \eta(\Omega) = e^{i\phi_p(\Omega)}[iu_p(\Omega)-i\text{sgn}(u_p(\Omega))\sqrt{u_p^2(\Omega)-1}].
\end{equation}

We can combine the results of Eqs. (\ref{eta_locked}-\ref{eta_unlocked}) by defining
\begin{equation}\label{eqn_Q}
    Q(u) = \bigg{\{}\mqty{iu + \sqrt{1-u^2} &\abs{u}\le 1 \\
    iu - i\text{sgn}(u)\sqrt{u^2-1} & \abs{u}\ge1},
\end{equation}
and write
\begin{equation}
    \eta(\Omega) = e^{i\phi_p(\Omega)}Q[u_p(\Omega)].
\end{equation}
Using this result, we obtain a self-consistency equation for $Z_p$,
\begin{equation}\label{eq_Z_int}
    Z_p(\Omega) = \int \abs{\Omega-\Omega'}^pg(\Omega')e^{i\phi_p(\Omega')}Q[\frac{\Omega' M_p}{K A_p(\Omega')}] \dd\Omega'.
\end{equation}

Equation (\ref{eq_Z_int}) depends only on the frequency distribution $g(\Omega)$, coupling exponent p and coupling strength $K$, and can be solved by conventional numerical methods (damped Picard fixed-point iteration\cite{kelley1995iterative,mastroianni2017nystrom,atkinson2001numerical} and additional details in Appendix C). We note that we assumed $g(\Omega)$ is symmetric and has zero mean, and that $K,p \ge0$.
The stationary order parameter can then be calculated using
\begin{equation}\label{eq_r_Z}
    \bar{r} = \abs{\int g(\Omega) \eta(\Omega) \dd\Omega}.
\end{equation}

\begin{figure}[h]
    \centering
    \includegraphics[trim= {0 0 0 0},clip,width=1\linewidth]{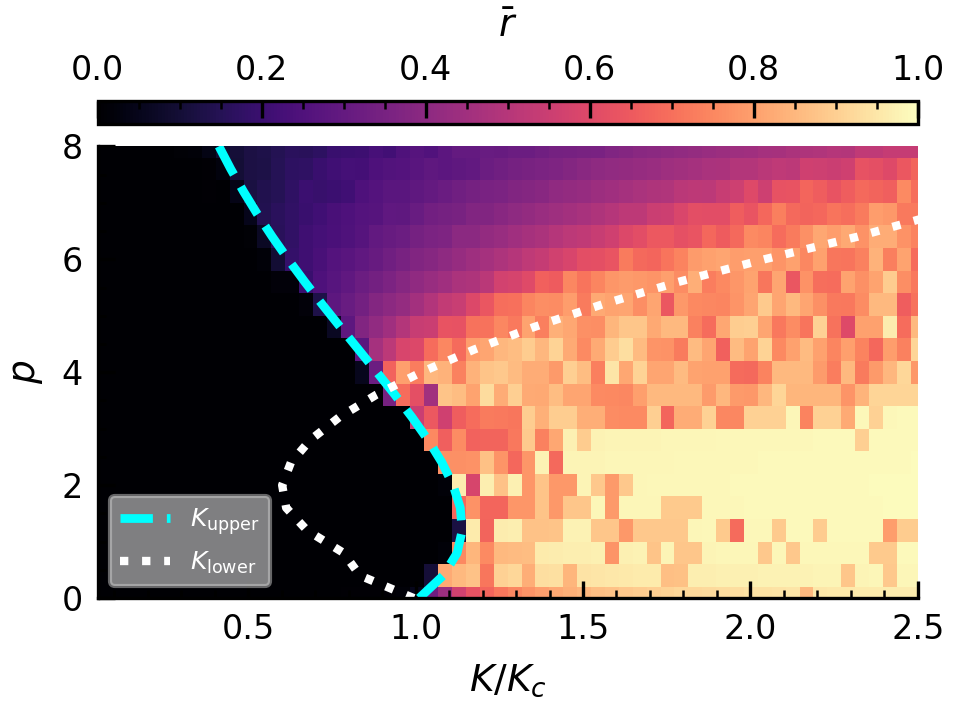}
    \caption{Semi-analytical calculation of the order parameter $\bar{r}$ as a function of normalized coupling strength $K/K_c$ and parameter $p$. $K_{upper},K_{lower}$ denotes the continuation limit of the unsynchronized and synchronized solutions respectively. When $K$ is increased above $K_{upper}$ (light blue dashed line), the unsynchronized state converges to synchronization. When $K$ is decreased below $K_{lower}$ (the white dotted line), the fully synchronized state converges to the unsynchronized state. The region where $K_{upper}>K_{lower}$ indicates the existence of a metastable synchronized state.}
    \label{fig_semianalytic}
\end{figure}
Figure \ref{fig_semianalytic} shows the stationary order parameter obtained from the self-consistency equation, Eq. (\ref{eq_r_Z}). The results are in good agreement with the numerical results presented in Fig. \ref{fig:power_2d}, including the abrupt synchronization transition in the intermediate-$p$ regime. 
We also verify the metastability of the synchronized state highlighted in Fig.  \ref{fig:hysteresis} by continuation of Eq. (\ref{eq_r_Z}). In the forward continuation, $K$ is increased incrementally and the converged solution at each value of $K$ is used as the initial condition at the next value. This procedure follows the unsynchronized branch until $K_{\mathrm{upper}}(p)$, where the solution leaves that branch and converges to a high synchronization solution. In the backward continuation, the calculation begins in the strongly synchronized regime and follows the synchronized branch as $K$ is decreased, until $K_{\mathrm{lower}}(p)$, below which the iteration converges to the low-coherence solution.

In the interval $K_{\mathrm{lower}}(p)<K<K_{\mathrm{upper}}(p)$ two different stationary self-consistent solutions are reached depending on the continuation direction, providing additional independent evidence for the observed hysteresis and metastability shown in Fig. \ref{fig:hysteresis}. Details regarding the evaluation of Eq. (\ref{eq_r_Z}) and $K_{upper,lower}$ can be found in Appendix C.

Lastly, it is interesting to consider why $p\approx 2$ appears to be optimal for synchronization. The coupling experienced by an oscillator with frequency $\Omega$ is
\begin{equation}\label{eq_eff_coup}
    \tilde{K}(\Omega) = \frac{K}{M_p}\int \abs{\Omega-\Omega'}^pg(\Omega')\dd\Omega'
\end{equation}

Assuming $g(\Omega)$ follows a normal distribution with standard deviation $\sigma$, we can write \cite{winkelbauer2012moments}
\begin{equation}
    M_p = \frac{(2\sigma)^p}{\sqrt{\pi}}\Gamma(\frac{p+1}{2}).
\end{equation}

For $\Omega\gg \sigma$, Eq. (\ref{eq_eff_coup}) yields
\begin{equation}\label{K_large_omega}
    \tilde{K}(\Omega)\approx \frac{K\sqrt{\pi}}{\Gamma(\frac{p+1}{2})}(\frac{\abs{\Omega}}{2\sigma})^p,
\end{equation}
in which case increasing $p$ can result in larger effective coupling felt by the oscillator. In comparison, for $\Omega \ll \sigma$,
\begin{equation}
    \tilde{K}(\Omega) \approx K \cdot 2^{-p/2},
\end{equation}
such that increasing $p$ results in less effective coupling felt by the oscillator. In other words, strongly detuned oscillators benefit from using more of the coupling budget as $p$ increases, while weakly detuned oscillators are left with exponentially less coupling. In section IV of the Supplemental Material we show analytically that in the strong coupling regime ($K\gg K_c, r\sim 1$), $1<p<2$ optimizes the budget distribution among oscillators in our scheme\cite{Supplemental}.

\section{Conclusions} We showed that under a fixed coupling budget, proportional coupling can improve synchronization in Kuramoto oscillator networks as compared to uniform (mean-field) coupling. With proportional coupling, an abrupt synchronization transition and hysteresis can occur, consistent with explosive synchronization. We have demonstrated that the improvement is robust to coupling disorder, and can be optimized for varying coupling budgets and sparse networks. Our results are robust to different frequency distributions, and apply to additional coupled oscillator models beyond the Kuramoto model.

Our investigations and results have both fundamental and practical implications. Proportional coupling schemes can benefit efforts to synchronize physical networks with minimal budgets in cases where the natural frequency distribution of the network is at least partially known or measurable, such as laser arrays. Fundamentally, our results reveal that the inclusion of a coupling budget constraint can alter the critical behavior of Kuramoto networks and highlights the relation between an abrupt synchronization transition and enhanced synchronization. Further investigations can adapt our approach to complex and sparse networks, different oscillator models, network topologies, and higher-order interactions.

\section*{Acknowledgments} 
A. P. and N. D. thank David Mukamel for valuable help and acknowledge support from the Minerva Stiftung, with funding from the Federal German Ministry for Education and Research. H. C. and N. V. thank Ying-Cheng Lai, Li-Li Ye, and Fan-Yi Lin for stimulating discussions and acknowledge funding support from the US Office of Naval Research under Grant No. N00014-24-1-2548.
\section*{Appendix}
\subsection{Numerical simulation methods}
To numerically simulate Kuramoto dynamics, we first randomly generate a frequency vector $\{\Omega_i\}$ using a specific frequency distribution, and normalize it such that its mean is $\ev{\Omega_i} = 0$ and its variance is $\ev{\Omega_i^2 - \ev{\Omega_i}^2} = \sigma^2$. The results in the main text were obtained using a normal frequency distribution, $g(\Omega)\sim \mathcal{N}(0,\sigma^2)$. We used $\sigma = 0.25$ for all simulations.

To generate the proportional coupling matrix, we use the generated frequency vector $\{\Omega_i\}$ to calculate the frequency difference matrix: $D_{ij}(p) = \abs{\Omega_i-\Omega_j}^p$ and normalize $D_{ij}$ according to Eq. (\ref{eqn_budget}).

We numerically simulate the dynamics of the network using an adaptive Runge-Kutta method of order 5(4) (RK45) and Eq. (\ref{kuramoto_eqn}). We run an adaptive timestep solver from $t\in\{0,T=100\}$ where the maximal step size is $dt=0.05$. The simulations are initiated with an initial condition of random phases. We evaluate the time averaged order parameter for each simulation realization 
$$\ev{r(T)}_\tau= \frac{1}{\tau}\int_{T-{\tau}}^T r(t)\dd{t},$$
where the averaging time $\tau = T/10$. We numerically validate that steady state has been reached by ensuring that the slope of the time averaged order parameter is small,
$$ \abs{\dv{\ev{r(t)}_\tau}{t}\Big{|}_{t=T}}<\delta,$$
where we take $\delta = 10^{-2}$. All simulations have converged with the simulation time.

The final averaged value of the order parameter, $\bar{r}$, is determined by averaging $r$ over 100 different realizations of $\{\Omega_i\}$ for each value of $K$ and 20 different initial conditions for each. For the inset of Fig. \ref{fig:prop_results}, 500 realizations were taken at each value of $K$. Our results are robust to changes in simulation time and size of the time step.

Numerical tests of the robustness of our results, specifically regarding apparent hysteresis can be found in Sec. VI of \cite{Supplemental}.

\subsection{Partially correlated proportional coupling}
For the results of Fig. (3), we generated two independent sets of frequency vectors, $\{\Omega_i\}$ and $\{\Omega^{\text{error}}_i\}$, and a coupling matrix
$$K_{ij}(\varepsilon) =\varepsilon D_{ij} + \sqrt{1-\varepsilon^2}D_{ij}^{\text{error}}$$

which ensures $\text{Corr}(K_{ij},D_{ij}) = \varepsilon$.
\subsection{Numerical evaluation of analytical results}
The evaluation of the order parameter $\bar{r}$ in Fig. \ref{fig_semianalytic} based on Eqs. (\ref{eq_Z_int}-\ref{eq_r_Z}) uses a fixed point iteration method \cite{kelley1995iterative,mastroianni2017nystrom,atkinson2001numerical}.

We generate a discrete sampling of $g(\Omega)$,
\[
\Omega_1,\Omega_2,\ldots,\Omega_{N_\Omega}
\]
and corresponding non-negative weights
\[
w_1,w_2,\ldots,w_{N_\Omega},
\qquad
\sum_{a=1}^{N_\Omega}w_a=1.
\]
such that
\[
\int_{-\infty}^{\infty}
g(\Omega)\,f(\Omega)\,d\Omega
\simeq
\sum_{a=1}^{N_\Omega}w_a h(\Omega_a),
\]
for general $f(\Omega)$.
We (randomly or deterministically) choose an initial $\eta(\Omega)$. We use our guess to evaluate $Z_p(\Omega)$ reevaluate $\eta(\Omega)$ through the self-consistency equations. All quantities of interest (such as $\eta(\Omega)$ and $Z_p(\Omega)$) are discretized on the same frequency grid as $g(\Omega)$.

At iteration $n$, $\eta(\Omega)$ is represented by
the complex vector
\[
\boldsymbol{\eta}^{(n)}
=
\left(
\eta_1^{(n)},\ldots,\eta_{N_\Omega}^{(n)}
\right),
\qquad
\eta_a^{(n)}\simeq\eta(\Omega_a).
\]
Such that
\[
Z_a^{(n)}
=
\sum_{b=1}^{N_\Omega}
w_b|\Omega_a-\Omega_b|^p
\eta_b^{(n)}.
\]
We define
\[
A_a^{(n)}=|Z_a^{(n)}|,
\qquad
\phi_a^{(n)}=\arg Z_a^{(n)},
\]

and calculate
\[
u_a^{(n)}
=
\frac{\Omega_aM_p}
{K A_a^{(n)}}.
\]
The stationary locked or drifting response associated with the current
field is
\[
\widehat{\eta}_a^{(n+1)}
=
e^{i\phi_a^{(n)}}Q\!\left(u_a^{(n)}\right),
\]
where $Q$ is defined in Eq. (\ref{eqn_Q}). 

To avoid oscillations between successive
estimates, we use damped Picard iteration,
\[
\eta_a^{(n+1)}
=
(1-\alpha)\eta_a^{(n)}
+
\alpha\widehat{\eta}_a^{(n+1)},
\]
where $0<\alpha\leq1$ is the damping parameter. 
Since solutions are invariant to global phase rotations, we remove the global phase at every iteration. 

Convergence is determined from the normalized residual
\[
\epsilon_n
=
\frac{
\abs{{\eta}^{(n+1)}
-
\boldsymbol{\eta}^{(n)}
}
}{
1+
\abs{{\eta}^{(n+1)}}
}.
\]
The iteration is considered converged when
\[
\epsilon_n<\epsilon_{FP}.
\]
A maximum of $2500$ iterations is allowed at each $(K,p)$ point, and nonconverged solutions are marked as such. 

After convergence, the stationary global order parameter is calculated
as
\[
r_{\mathrm{st}}
=
\left|
\sum_{a=1}^{N_\Omega}
w_a\eta_a
\right|.
\]

A full table of parameters used in the process can be found in Sec. VI of \cite{Supplemental}.

To calculate $K_{upper},K_{lower}$, we follow the stationary branches as the coupling is ramped up or down. For the forward continuation, we begin the process with $K=0$ and set the initial $ \eta(\Omega) = 10^{-4}$. After convergence, we ramp $K^{(m+1)} = K^{(m)}+\Delta K$, and use the converged $\eta(\Omega)$ as the initial condition. For the backward continuation, we begin with $K=K_{max}$ and $\eta(\Omega)=1$, and follow the same procedure in the opposite direction. We identify $K_{upper},K_{lower}$ by the largest jump in the value of the order parameter along the trajectory.

\FloatBarrier

\clearpage
\newpage
\widetext
\begin{center}
\textbf{\Large Supplemental Materials}
\end{center}
\setcounter{figure}{0}
\renewcommand{\figurename}{Fig.}
\renewcommand{\thefigure}{S\arabic{figure}}
\setcounter{section}{0}
\section{Numerical validation of hysteresis}

To numerically study hysteresis in the Kuramoto simulation, we employ a similar numerical method to that described in appendix A, while adiabatically ramping the coupling up or down. We start with $K=0$ and some realization of disorder $\{\Omega\}$ and random initial conditions. We run the simulation until reaching steady state. We then run a new simulation with the same realization of disorder, but with $K$ increased by $\Delta K/K_c=0.05$ and with the initial condition given by the final state of phases in the preceding simulation. We repeat this process until we reach the maximal value of $K$, at which point we start reducing $K$ in each simulation. The entire process is reiterated multiple times for different realizations of disorder.

\begin{figure}[H]
    \centering
    \includegraphics[width=0.75\linewidth]{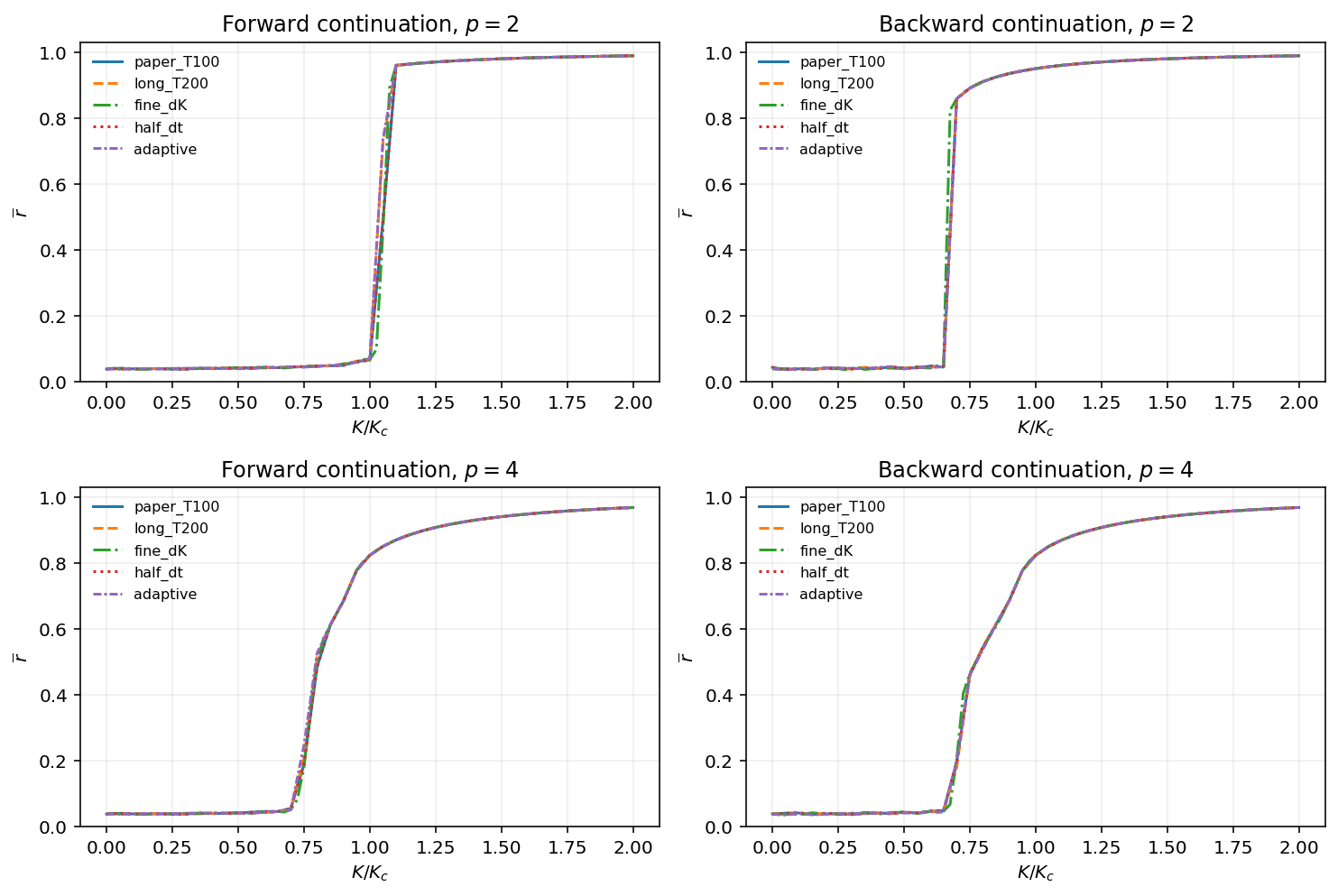}
    \caption{Numerical tests of the robustness of the hysteresis phenomenon, showing $\bar{r}(K)$ for different numerical simulation parameters. The simulations vary in the following ways: \textit{paper\_T100} - the parameters used in the simulations in the main text; \textit{long\_T200} - double total simulation time; \textit{fine\_dK} - halved coupling magnitude increment; \textit{half\_dt} - halved solver time step size; \textit{adaptive} - adaptive step size and integration time according to numerical convergence. }
    \label{fig_sup_hyst}
\end{figure}

To validate the robustness of the hysteresis results to numerical simulation details, we repeated the simulation for various choices of total integration time $T$, time step size $dt$ and coupling step size $\Delta K$ for multiple initial conditions. The results are presented in Fig. \ref{fig_sup_hyst}, and show that the observed hysteresis appears robust to numerical simulation details.

\section{Numerical results for other frequency detuning distributions}
\begin{figure}[H]
    \centering
    \includegraphics[width=0.5\linewidth]{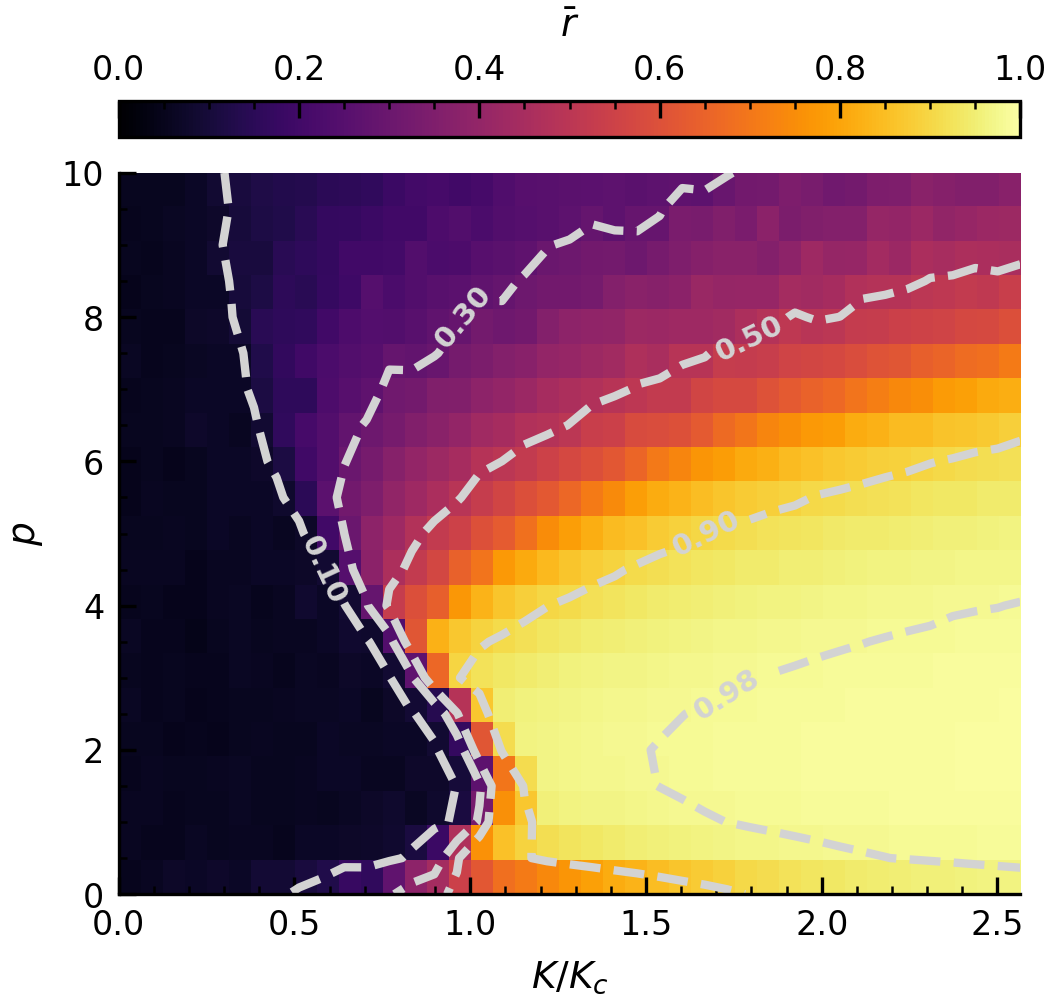}
    \caption{Synchronization of a network of $N = 512$ oscillators for $K_{ij}\propto\abs{\Omega_i-\Omega_j}^p$, where $g(\Omega)$ is a Cauchy distribution.}
    \label{fig_S_cauchy}
\end{figure}

\begin{figure}[H]
    \centering
    \includegraphics[width=0.5\linewidth]{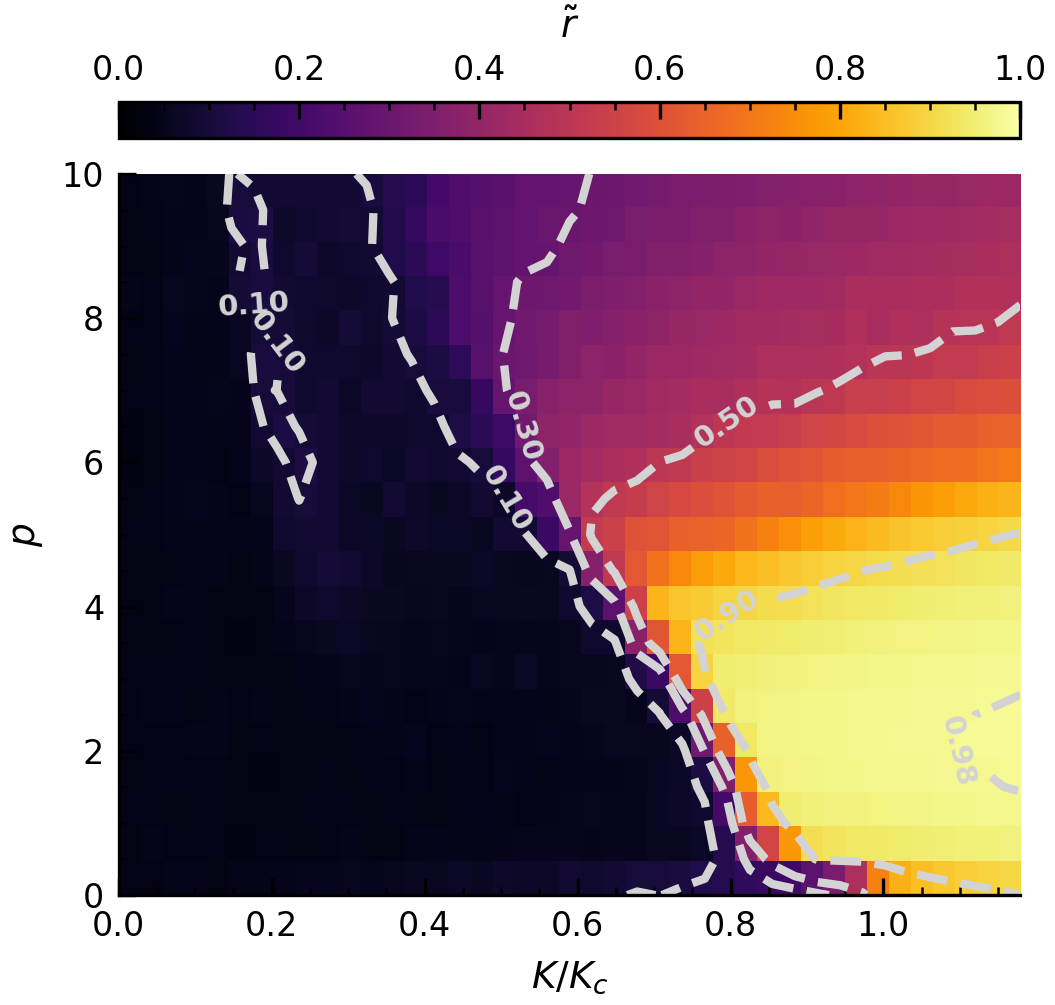}
    \caption{Synchronization of a network of $N = 512$ oscillators for $K_{ij}\propto\abs{\Omega_i-\Omega_j}^p$, where $g(\Omega)$ is a uniform distribution.}
    \label{fig_S_uniform}
\end{figure}

We repeated the simulations for other distributions of frequency detuning, $g(\Omega)$ using the same numerical protocol described above. Figures \ref{fig_S_cauchy}-\ref{fig_S_uniform} show the results for a Cauchy distribution and uniform frequency distributions. The results reveal a strong similarity to the results shown in the main paper, indicating that the synchronization enhancement effect is robust to changes in the frequency distribution.

\section{Proportional coupling in Stuart-Landau oscillators}
To test whether the synchronization enhancement obtained with proportional coupling extends beyond the Kuramoto model, we simulated a network of \(N=64\) Stuart–Landau oscillators. The Stuart–Landau oscillator is the universal amplitude equation near a supercritical Hopf bifurcation and extends a phase-only model by retaining amplitude dynamics\cite{matthews1991dynamics}. The dynamics of oscillator \(i\), described by the complex amplitude \(z_i\), are
\[
\dot z_i=\left(1+i\omega_i-|z_i|^2\right)z_i
+\sum_{j=1}^{N}K_{ij}(z_j-z_i),
\]

The intrinsic frequencies were drawn independently from a normal distribution with mean \(\bar\omega=1\) and standard deviation \(\sigma_\omega=0.15\); as in the Kuramoto simulations, each finite realization was subsequently shifted and rescaled to have exactly the prescribed sample mean and standard deviation. The coupling coefficients were chosen according to the same generalized proportional-coupling rule as in the main text,
\[
K_{ij}
=
K\frac{|\omega_i-\omega_j|^p}{C_p},
\qquad
C_p=\frac{1}{N}\sum_{m,n}|\omega_m-\omega_n|^p,
\]
such that \(N^{-1}\sum_{ij}K_{ij}=K\) for every disorder realization and every value of \(p\). Initial conditions were independently randomized for both phase and amplitude, with \(\arg z_i(0)\) uniformly distributed on \([0,2\pi)\) and \(|z_i(0)|\) uniformly distributed on \([0.7,1.3]\). The coupled equations were integrated using a fourth-order Runge–Kutta scheme with time step \(dt=0.02\) up to \(T=100\). Synchronization was quantified by the instantaneous phase order parameter
\[
R(t)=\left|\frac{1}{N}\sum_{j=1}^{N}
e^{i\arg z_j(t)}\right|,
\]
and the steady-state value for each simulation was obtained by averaging \(r(t)\) over the final \(10\%\) of the integration interval. For each pair \((K,p)\), the reported order parameter was averaged over \(30\) independent realizations of the frequency disorder and \(20\) independent initial conditions per realization. 

\begin{figure}[H]
    \centering
    \includegraphics[width=0.75\linewidth]{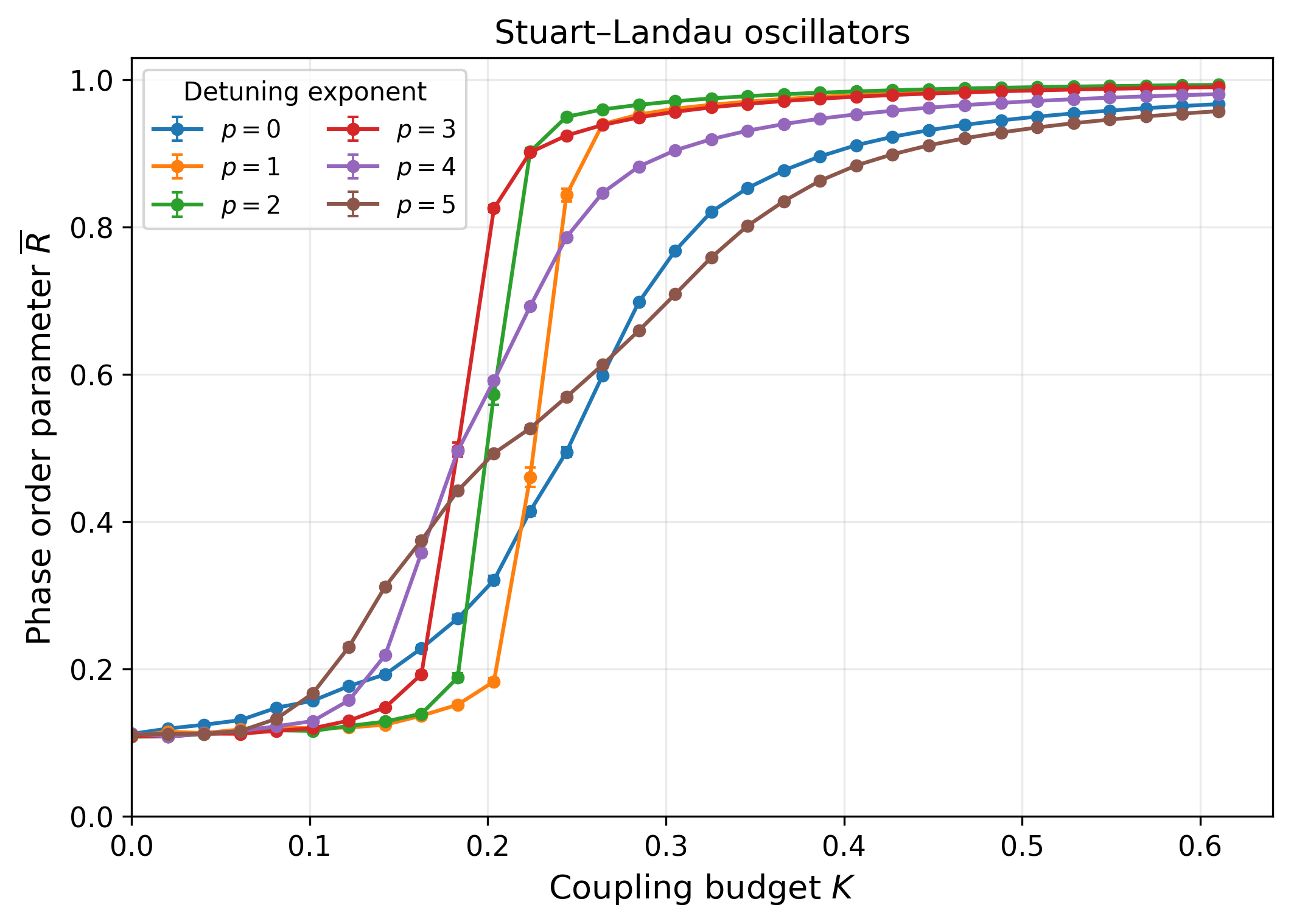}
    \caption{Synchronization order parameter $r$ as a function of coupling budget for Stuart-Landau oscillators with proportional coupling of the form $K_{ij}\propto |\omega_i-\omega_j|^p$. Error bars represent standard deviation of the order parameter across random frequency detuning distribution realizations. }
    \label{fig_SL}
\end{figure}

The results are presented in Fig. \ref{fig_SL} and show that intermediate values of $p$ enhance phase synchronization relative to uniform coupling.

\section{Analytics in the strong synchronization regime}
\subsection{Deriving the existence condition of the synchronized state}
The Kuramoto model equation for oscillator i is
\begin{equation}
  \dot{\theta_i} = \Omega_i + \sum_{j}K_{ij}\sin(\theta_j-\theta_i)  
\end{equation}

For strong coupling ($K\gg K_c$) and $r\sim 1$, we can make the following approximations:
\begin{enumerate}
    \item All oscillators are locked, $\dot{\theta}_i=0$.
    \item $r\sim 1$ implies that all $\theta_i$ are close to the mean value, $$\abs{\theta_i-\theta_j}\ll1$$
    so
    $$\sin(\theta_j-\theta_i)\approx \theta_j-\theta_i$$
    \item The coupling matrix and frequency distribution are symmetric under $\Omega_i\leftrightarrow -\Omega_i$, implying that oscillators with opposite detuning will settle on opposite phases:
    $$\theta(\Omega) = - \theta(-\Omega)$$

\end{enumerate}

Using these assumptions yields
$$\Omega_i - \theta_i \sum_j K_{ij} + \sum_jK_{ij}\theta_j = 0$$

We define the effective frequency detuning and coupling:
\begin{equation}\label{eqn_omega_eff}
  \tilde{\Omega}(\Omega_i) \equiv \Omega_i + \sum_j K_{ij}\theta_j  
\end{equation}
\begin{equation}\label{eqn_K_eff}
  \tilde{K}(\Omega_i) \equiv \sum_j K_{ij}  
\end{equation}

Such that
\begin{equation}\label{eqn_condition}
  \theta_i = \frac{\tilde{\Omega}(\Omega_i)}{\tilde{K}(\Omega_i)} \ll 1  
\end{equation}

is an effective self-consistency condition. The synchronized solution exists if condition (\ref{eqn_condition}) is satisfied for all $i$.

In the next sections, we study the limiting behavior of $\tilde{K}$ and $\tilde{\Omega}$.
\subsection{The effective coupling}
In the continuum limit ($N \rightarrow \infty$) the effective coupling can be rewritten as
\begin{equation}\label{eq_nom}
\tilde{K}(\Omega_i) = \sum_j K_{ij} \rightarrow \frac{1}{M_p} K\int_{-\infty}^{\infty}\abs{\Omega-\Omega'}^p\delta(\Omega-\Omega_i)g(\Omega)g(\Omega')\dd\Omega'\dd\Omega = \frac{1}{M_p}K \ev{\abs{\Omega_i-\Omega}^p}
\end{equation}

For $p>-1$, this is the absolute $p$-th moment of the frequency distribution shifted around $\Omega_i$, where
$M_p$ is the normalization constant
\begin{equation}\label{eq_denom}
    M_p \equiv \iint \abs{\Omega-\Omega'}^p g(\Omega)g(\Omega')\dd\Omega\dd\Omega' \equiv\ev{\abs{\Delta\Omega}^p}  
\end{equation}

For normal distribution, we can get analytical expressions as function of p.
If $\Omega_i \in \mathcal{N}(0,\sigma^2)$, then $\Omega_i-\Omega_j \in \mathcal{N}(0,2\sigma^2)$. Using known results for normal distribution \cite{winkelbauer2012moments}, Eq. (\ref{eq_nom}) yields 
\begin{equation}
    \ev{\abs{\Delta\Omega}^p} = \frac{1}{\sqrt{\pi}}(2\sigma)^p \Gamma(\frac{p+1}{2})
\end{equation}

An explicit exact solution for Eq. (\ref{eq_denom}) does not exist, so we consider a solution at two limits: $\abs{\Omega_i}\ll\sigma$, and $\abs{\Omega_i}\gg \sigma $.

For $\abs{\Omega_i}\ll \sigma$,
\begin{equation}
    \ev{\abs{\Omega_i-\Omega}^p}\approx \ev{\abs{\Omega}^p} =\frac{1}{\sqrt{\pi}}\sigma^p 2^{p/2} \Gamma(\frac{p+1}{2})
\end{equation}

for which the effective coupling is,
\begin{equation}\label{K_small_detuning}
    \tilde{K}(\Omega_i) = \frac{K\ev{\abs{\Omega_i-\Omega}^p}}{\ev{\abs{\Delta\Omega}^p}} \approx K \cdot 2 ^{-p/2}
\end{equation}

At this limit the coupling is independent of $\Omega_i$ but decays exponentially with $p$.
For the case of $\abs{\Omega_i}\gg \sigma$,
\begin{equation}
    \ev{\abs{\Omega_i-\Omega}^p} \approx \ev{\abs{\Omega_i}^p}\approx \abs{\Omega_i}^p
\end{equation}

for which the coupling is
\begin{equation}\label{eqn_K_large_detuning}
    \tilde{K}(\Omega_i) = \frac{K\sqrt{\pi}}{\Gamma(\frac{p+1}{2})}(\frac{\Omega_i}{2\sigma})^p
\end{equation}

At this limit the coupling scales as $\abs{\Omega_i}^{p}$. For $p>0$, oscillators with larger detuning are easier to lock, while oscillators with smaller detuning are harder to lock.

Figure \ref{fig:analysis_subplots}(a) shows a numerical evaluation of $\tilde{K}$ as a function of $\Omega_i$ for different values of $p$.

\subsection{The effective detuning}
The locking is set by the ratio of the effective coupling to the effective detuning. The effective detuning in the continuum limit is
\begin{equation}\label{eqn_eff_det}
   \tilde{\Omega}(\Omega_i) = \Omega_i + \frac{K}{\ev{\abs{\Delta\Omega}^p}} \int_{-\infty}^{\infty} \abs{\Omega_i-\Omega}^p \theta(\Omega) g(\Omega)\dd\Omega 
\end{equation}

Equation (\ref{eqn_eff_det}) can be rewritten without an explicit dependence on $\theta$ by using Eq.
(\ref{eqn_condition}) to yield

\begin{equation}\label{eq_integral_eqn}
    \tilde{\Omega}(\Omega_i) = \Omega_i +\frac{K}{\ev{\abs{\Delta\Omega}^p}}\int_{-\infty}^{\infty}\abs{\Omega_i-\Omega'}^p\frac{\tilde{\Omega}(\Omega')}{\tilde{K}(\Omega')}g(\Omega')\dd\Omega' = \Omega_i +\int_{-\infty}^{\infty}\abs{\Omega_i-\Omega'}^p\frac{\tilde{\Omega}(\Omega')}{\ev{\abs{\Omega'-\Omega}^p}}g(\Omega')\dd\Omega'
\end{equation}

Note that Eq. (\ref{eq_integral_eqn}) does not depend on $K$.

Finally, for $p=2$, $\tilde{\Omega}(\Omega_i) \propto \Omega_i$, such that it renormalizes the frequency distribution. Equation (\ref{eqn_eff_det}) reduced to
\begin{equation}
    \tilde{\Omega}(\Omega_i) = \Omega_i+\frac{K}{\ev{\abs{\Delta\Omega}^2}} \int_{-\infty}^{\infty}(\Omega_i-\Omega)^2\theta(\Omega)g(\Omega)\dd\Omega
\end{equation}

The integral is taken over an even domain. Since $g(\Omega)$ is even and $\theta(\Omega)$ is odd, the integrand yields a non-vanishing contribution only for terms with odd powers of $\Omega$, so
\begin{equation}
    \tilde{\Omega}(\Omega_i) = \Omega_i-2\frac{K}{\ev{\abs{\Delta\Omega}^2}}\int_{-\infty}^{\infty}\Omega_i\Omega\theta(\Omega)g(\Omega)\dd\Omega \equiv \Omega_i (1-\beta)
\end{equation}
where
\begin{equation}\label{eqn_beta}
  \beta \equiv 2\frac{K}{\ev{\abs{\Delta\Omega}^2}}\int_{-\infty}^{\infty}\Omega\theta(\Omega)g(\Omega)\dd\Omega  
\end{equation}

We can solve Eq. (\ref{eqn_beta}) by using the synchronization condition:
\begin{equation}
    \theta_i = \frac{\tilde{\Omega}(\Omega_i)}{\tilde{K}(\Omega_i)}=\frac{\Omega_i(1-\beta)}{\tilde{K}(\Omega_i)}
\end{equation}

to yield
\begin{equation}
    \beta = 2\frac{K}{\ev{\abs{\Delta\Omega}^2}}\int_{-\infty}^{\infty}\Omega \frac{\Omega(1-\beta)}{\tilde{K}(\Omega)}g(\Omega)\dd\Omega =  (1-\beta) \cdot \int_{-\infty}^{\infty}\frac{2\Omega'^2}{\ev{\abs{\Omega'-\Omega}^2}}g(\Omega')\dd\Omega'
\end{equation}

This value of $\beta$ is given by
\begin{equation}
   \beta = \frac{\mathcal{I}}{1+\mathcal{I}} \rightarrow  \mathcal{I}\equiv \int_{-\infty}^{\infty}\frac{2\Omega^2}{\ev{\abs{\Omega-\Omega'}^2}
}g(\Omega)\dd\Omega
\end{equation}

and was numerically evaluated as $\beta \approx 0.4$.

Figure \ref{fig:analysis_subplots}(b) shows a numerical evaluation of $\tilde{\Omega}$ as a function of $\Omega_i$ for different values of $p$. The results reveal that the effective frequency is determined by the frequency distribution and is independent of the coupling strength. For $p=2$ a linear correction reduces the effective frequency of all oscillators.
For $p<2$ the frequency correction is sublinear such that high frequencies have less correction. For $p>2$, $\tilde{\Omega}$ is a higher-order polynomial in $\Omega_i$. As a result, $\tilde{\Omega_i}$ can have the opposite sign of $\Omega_i$ for large enough frequencies. 

\subsection{The synchronization condition}

As noted earlier, the existence of synchronization is determined by Eq. (\ref{eqn_condition}). We evaluate Eq. (\ref{eqn_condition}) for different values of $p$ using the expressions derived above for $\tilde{K}$ and $\tilde{\Omega}$.

For $p=0$, we get
\begin{equation}\label{eqn_p_0}
\frac{\abs{\Omega_i}}{K}\ll 1
\end{equation}

Equation (\ref{eqn_p_0}) indicates that for unbounded frequency distributions (such as the normal distribution), there will always exist some oscillators that are sufficiently detuned so they will not synchronize.

For large $p$, Eqs. (\ref{eqn_K_large_detuning}-\ref{eqn_eff_det}) show that for large detunings ($\abs{\Omega_i}\gg \sigma$),
\begin{equation}
    \theta_i \propto \abs{\Omega_i}^{-1}
\end{equation}
will always be smaller than $1$. However, near the center of the distribution ($\abs{\Omega_i}\ll\sigma$),
\begin{equation}
  \theta_i \sim \frac{2^{p/2}\abs{\Omega_i}}{K}
\end{equation}

For nonzero $\Omega_i$, the exponential prefactors can overcome the small detuning of weakly detuned oscillators  ($\abs{\Omega_i}\lesssim\sigma$) and break synchronization. The critical value of $\Omega_i$ is reduced as $p$ increases.

For $1<p<2$, these two effects balance out: A linear correction to the frequencies keeps the effective detuning smaller than the natural detuning, ($|{\tilde{\Omega}(\Omega_i)}| < \abs{\Omega_i}$), while the coupling still increases for strongly detuned oscillators. For $p=2$, there exists a closed form solution
\begin{equation}
    \theta_i = \frac{2\sigma^2\Omega_i(1-\beta)}{K(\sigma^2+\abs{\Omega_i}^2)}\ll1 
\end{equation}

which is bounded and maximal at $\abs{\Omega_i}=\sigma$. The synchronized solution exists if $K$ is sufficiently larger than $\sim 0.6\sigma$.

\begin{figure}
    \centering
    \includegraphics[width=1\linewidth]{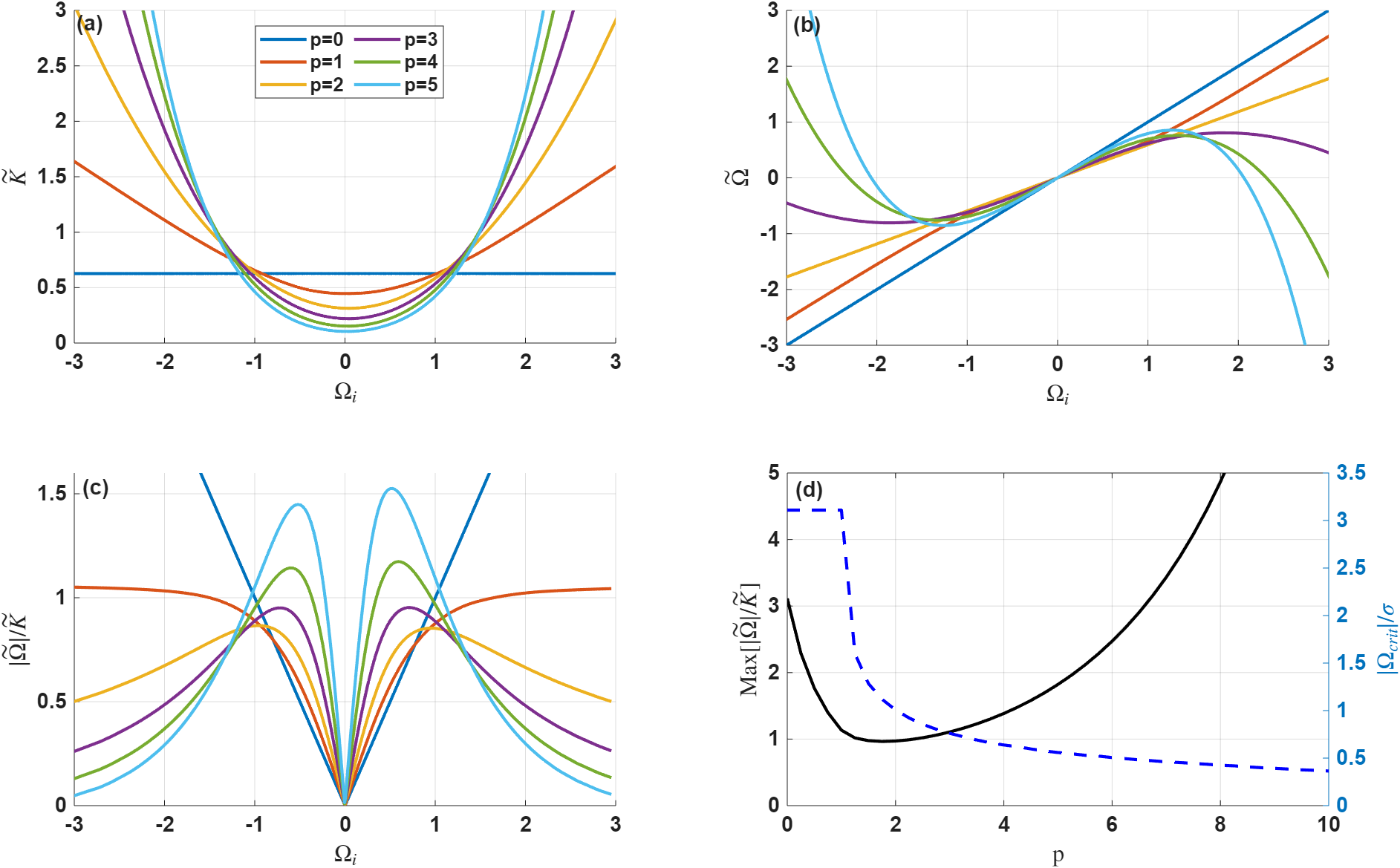}
    \caption{Numerical evaluation of the synchronization condition (Eq. (\ref{eqn_condition})). \textbf{(a):} The effective coupling $\tilde{K}$ as a function of detuning $\Omega_i$. \textbf{(b):}The effective detuning $\tilde{\Omega}$ as a function of detuning $\Omega_i$. \textbf{(c):} $\tilde{\Omega}/\tilde{K}$  as a function of $\Omega_i$. \textbf{(d):} The maximal value of $\tilde{\Omega}/\tilde{K}$ as a function of $p$ (solid black line), and the value of $\Omega_i$ for which it is obtained (dashed blue line). }
    \label{fig:analysis_subplots}
\end{figure}

Figure \ref{fig:analysis_subplots}(c) shows a numerical evaluation of Eq. (\ref{eqn_condition}) for different values of $\Omega_i$ and $p$. As evident, for $p=0$ the ratio $\tilde{\Omega}/\tilde{K}$ diverges linearly with $\abs{\Omega_i}$ as expected. This divergence decreases as $p$ is increased, and the ratio is bounded and can be found in closed form for $p=2$. For $p>2$, the maximal value of the ratio increases for small $\Omega_i$. Figure \ref{fig:analysis_subplots}(d) shows the maximal value of $\tilde{\Omega}/\tilde{K}$ for different values of $p$. As evident, the lowest value can be found around $p\approx 2$. Figure \ref{fig:analysis_subplots} also shows that the $\Omega_i$ for which Eq. (\ref{eqn_condition}) is maximal decreases as $p$ increases.

\section{Frequency distribution of oscillators}
\begin{figure}[H]
    \centering
    \includegraphics[width=0.8\linewidth]{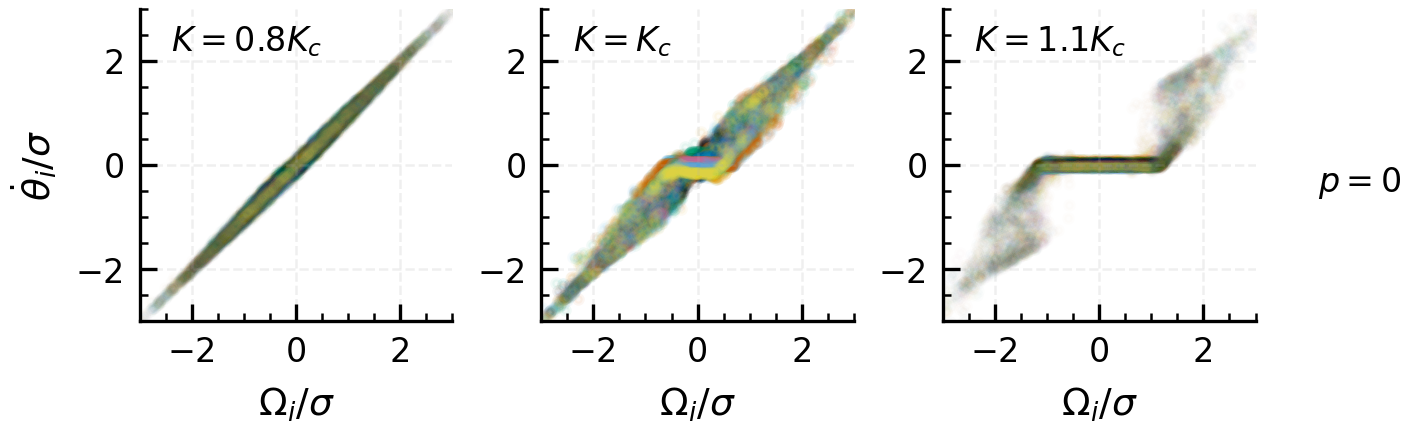}
    \includegraphics[width=0.8\linewidth]{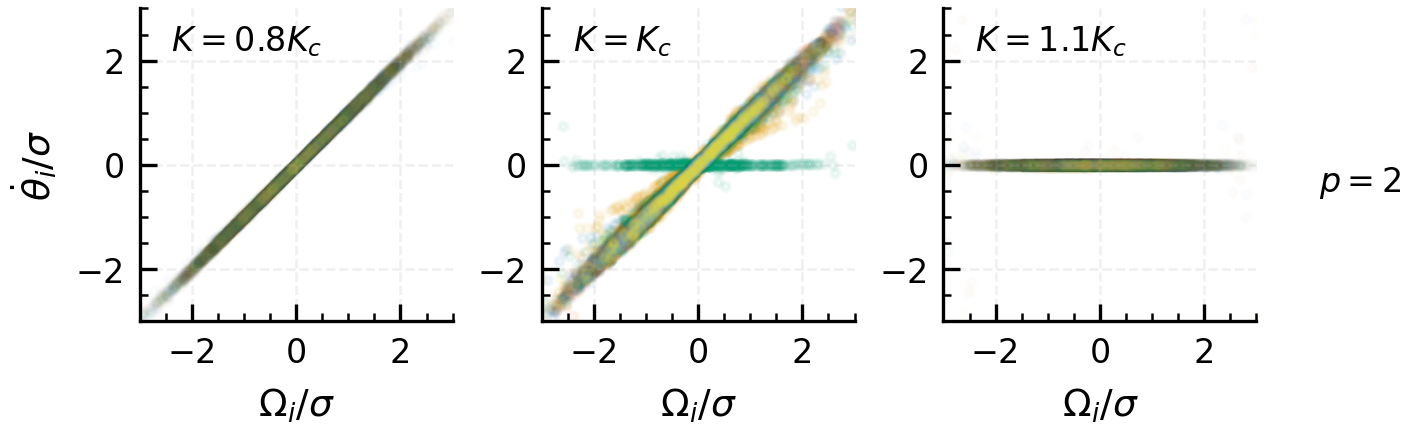}
    \includegraphics[width=0.8\linewidth]{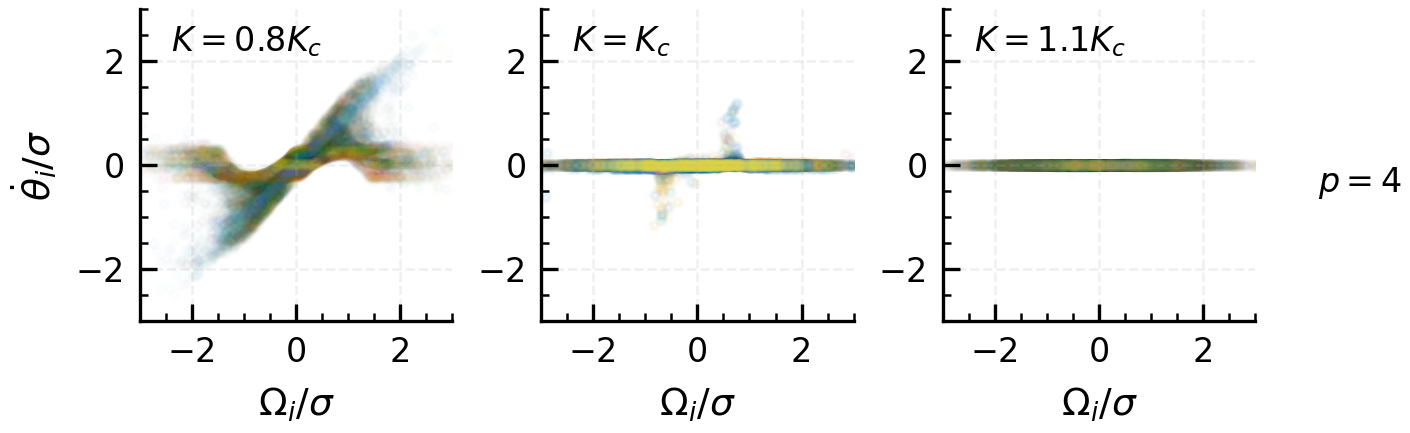}
    \caption{Steady state frequency $\dot{\theta}_i$ as function of initial detuning $\Omega_i$ for different $p$ and coupling strength $K/K_c$. Top, middle and bottom rows correspond to $p=0,2,4$ respectively. Left, middle, and right columns correspond to $K=0.8,1,1.1K_c$ respectively. }
    \label{fig_in_out}
\end{figure}

We consider the steady state frequency of the different oscillators $\dot{\theta}_i$ as a function of detuning $\Omega_i$.  The results are presented in Fig. \ref{fig_in_out}, and are  obtained from the same simulations presented in the main text. The rows of Fig. \ref{fig_in_out} correspond to different values of $p$, and the columns correspond to different values of $K/K_c$. The results obtained from different realizations of the frequency distribution are shown on the same plot in different colors. 

For $p=0$ (uniform all-to-all coupling), increasing  $K$ leads to a larger synchronized cluster, evident by an increasing plateau in the frequency distribution at $\dot{\theta}_i=0$. The first oscillators to synchronize are those with $\abs{\Omega_i}\ll \sigma$, and oscillators with larger detuning synchronize as the coupling is increased. Far detuned oscillators are unable to synchronize even for large coupling values.
For $p=2$, an abrupt transition is observed around $K=K_c$. As evident, some realizations show a completely unsynchronized frequency distribution ($\dot{\theta}_i\propto\Omega_i)$, while some show a completely synchronized distribution ($\dot{\theta_i}=0$). As $K$ exceeds $1.1K_c$, the network successfully synchronizes all oscillators for all disorder realizations.
For $p=4$, far detuned oscillators begin to synchronize even when $K<K_c$, but oscillators with $\Omega_i\approx\pm\sigma$ fail. In fact, it is evident that these oscillators are the last to synchronize, around $K=K_c$.

\newpage
\section{Table of numerical parameters for evaluation of analytical results}
\begin{table*}[h]
\centering
\caption{\label{tab:fixed_point_parameters}
Numerical parameters used to solve Eq. (20). The coupling grid is expressed in units of the standard
Kuramoto critical coupling $K_c=2/[\pi g(0)]$.}
\renewcommand{\arraystretch}{1.18}
\setlength{\tabcolsep}{6pt}
\begin{tabular}{p{0.22\textwidth}p{0.16\textwidth}p{0.53\textwidth}}
\hline\hline
Quantity & Value & Comment \\
\hline
Frequency distribution
& $\mathcal{N}(0,1)$
&  \\

Frequency truncation
& $|\Omega|\leq4.5\sigma$
& Distribution is renormalized after truncation. \\

Number of frequency nodes
& $N_\Omega=128$
& \\

Exponent grid
& $0\leq p\leq8, \Delta p =0.25$
&  \\

Coupling grid
& $0\leq K/K_c\leq2.50, \Delta K / K_c \simeq 0.0415$
&  \\

Damping parameter
& $\alpha=0.40$
&  \\

Convergence tolerance
& $\epsilon_{\mathrm{FP}}=10^{-9}$
& \\

Maximum iteration number
& $n_{\mathrm{iter}}^{\max}=2500$
&  \\

Continuation perturbation
& $\epsilon_{\mathrm{seed}}=10^{-4}$
& Small complex perturbation added during continuation to avoid
artificial trapping at the exactly incoherent fixed point. \\

Forward initial condition
& Random phases
& The initial frequency-resolved responses have unit magnitude and
independent phases uniformly distributed on $[0,2\pi)$. \\

Backward initial condition
& $\eta(\Omega)=1$
& The backward continuation begins from a fully synchronized state
 at the largest coupling. \\

\hline\hline
\end{tabular}
\end{table*}
\bibliography{main}

\end{document}